\documentclass[preprint,12pt,review]{elsarticle}

\usepackage[boxruled]{algorithm2e}
\usepackage{subfigure}
\usepackage{multirow}
\usepackage{framed}
\usepackage{amssymb}
\usepackage{amsmath}
\usepackage{hyperref}
\usepackage{bm}
\usepackage{xcolor}
\usepackage[margin=1in]{geometry}

\journal{Computer Networks}

\newcommand{\name}{{\sf Iris}}

\begin{document}

\begin{frontmatter}

\title{Efficient Inter-Datacenter Bulk Transfers with Mixed Completion Time Objectives\tnoteref{t1}}

\tnotetext[t1]{The work is partially supported by a Cisco Faculty Research Award under grant number 1312182.}

\author{Mohammad Noormohammadpour}
\address{University of Southern California}
\author{Srikanth Kandula}
\address{Microsoft}
\author{Cauligi S. Raghavendra}
\address{University of Southern California}
\author{Sriram Rao}
\address{Facebook}

\begin{abstract}
Bulk transfers from one to multiple datacenters can have many different completion time objectives ranging from quickly replicating some $k$ copies to minimizing the time by which the last destination receives a full replica. We design an SDN-style wide-area traffic scheduler that optimizes different completion time objectives for various requests. The scheduler builds, for each bulk transfer, one or more multicast forwarding trees which preferentially use lightly loaded network links. Multiple multicast trees are used per bulk transfer to insulate destinations that have higher available bandwidth and can hence finish quickly from congested destinations. These decisions--how many trees to construct and which receivers to serve using a given tree--result from an optimization problem that minimizes a weighted sum of transfers' completion time objectives and their bandwidth consumption. Results from simulations and emulations on Mininet show that our scheduler, {\name}, can improve different completion time objectives by about $2.5\times$.
\end{abstract}





\end{frontmatter}

\section{Introduction} \label{introduction}
A wide range of distributed applications replicate content and data to increase end-users' quality of experience \cite{configurator, facebook-express-backbone, netflix-replication, b4, blast, replica_far, wassermann2017anycast, huici2017efficient} which results in inter-datacenter bulk multicast transfers with a given set of receivers. For a variety of applications, objects may be replicated to at least four datacenters and for some applications potentially to tens of datacenters \cite{dccast, cdn_survey}. Moreover, an analysis of Baidu's traffic~\cite{overlay_hkust} across $30$ datacenters showed that over $90\%$ of the traffic is multicast and over $90\%$ of the multicast transfers are destined to at least $60\%$ of datacenters. 

A variety of approaches can be used to perform bulk multicast transfers. We can model a bulk multicast transfer as multiple independent unicast bulk transfers \cite{tempus, owan, swan, b4} which wastes network capacity and can increase the transfer completion times. Standard internet multicasting \cite{ip_multicast} builds multicast trees incrementally as receivers join the multicast session without considering the distribution of traffic load across network links. Therefore, generated multicast trees can be considerably larger than necessary with highly unbalanced network load distribution. Overlay multicasting \cite{nice} builds application layer multicast trees which may be far from the optimal due to limited visibility of network link level status and little control of traffic routing at the network layer. Peer-to-peer file distribution \cite{slurpie, bittorrent} aims to maximize throughput per receiver in a decentralized fashion and greedily, which can be far from global optimization. Centralized multicast routing approaches allow for better multicast tree selection by incorporating a global view of network status. Some centralized methods, such as \cite{avalanche, datacast}, target the regular and structured topologies of networks inside datacenters, which cannot be directly applied to inter-datacenter networks. Many other centralized techniques, such as \cite{raera, sdn_multicast, MPMC_2013, MPMC_2016, ddccast, AGE, dartree}, do not consider the optimization of receiver completion times, especially in an online scenario with many concurrent bulk multicast transfers. Finally, very recently, several centralized proposals aim to optimize the completion times of bulk multicast receivers \cite{dccast, quickcast}. Our work in this paper builds on these techniques.

When receivers of a bulk multicast transfer have very different network bandwidth available on paths from the sender, the slowest receiver dictates the completion time for all receivers. Recent work suggests using multiple multicast trees to separate the faster receivers which will improve the average receiver's completion time~\cite{quickcast}. However, each additional tree consumes more network bandwidth and at the extremum, this idea devolves to one tree per receiver. We aim to answer the following questions:

\begin{enumerate}
    \item What is the right number of trees per transfer?
    \item Which receivers should be grouped in each tree?
\end{enumerate}

We analyze a relaxed version of this partitioning problem where each partition is a subset of receivers attached to the sender with a separate forwarding tree. We first propose a partitioning technique that reduces the average receiver completion times of receivers by isolating slow and fast receivers. We study this approach in the relaxed setting of having a congestion-free network core, i.e., links in/out of the datacenters are the capacity bottlenecks, and considering max-min fair rate allocation from the underlying network. We then develop a partitioning technique for real-world inter-datacenter networks, without relaxations, and inspired by the findings from studying the relaxed scenario. The partitioning technique operates by building a hierarchy of valid partitioning solutions and selecting the one that offers the best average receiver completion times. Our evaluation of this partitioning technique on real-world topologies, including ones with bottlenecks in the network core, show that the technique yields completion times that are close to a lower bound and hence nearly optimal.

Back-end geo-distributed applications running on datacenters can have different requirements on how their objects are replicated to other datacenters. Hence, inter-datacenter traffic is usually a mix of transfers with various completion time objectives. For example, while reproducing $n$ copies of an object to $n$ different datacenters/locations, one application may want to transfer $k$ copies quickly to any among $n$ given receivers, and another application may want to minimize the time when the last copy finishes. In the former case, grouping the slower $n-k$ receivers into one partition consumes less bandwidth and this spare bandwidth could be used to speed up the other transfers. In the latter case, by grouping all receivers except the slowest receiver together to use one tree, we can isolate them from the slowest receiver with minimal bandwidth consumption. Minimizing the completion times of all receivers is another possible objective. Our technique takes as input a binary objective vector whose $i$\textsuperscript{th} element expresses interest in the completion time of the $i$\textsuperscript{th} fastest receiver; it aims to minimize the completion times of receivers whose rank is set to one in this objective vector. It is easy to see that following values of the objective vector achieve the goals discussed so far; when $k=1$, $n=3$, $\{1, 0, 0\}$, $\{0, 0, 1\}$ and $\{1, 1, 1\}$ aim to minimize the completion time of the fastest $k$ out of $n$ receivers, the slowest receiver, and all receivers, respectively.

We have built a system called {\name} which combines the proposed partitioning technique and the application/user supplied objective vectors. It operates in a logically centralized manner, receives bulk multicast transfer requests from end-points, and computes receiver partitions along with their multicast forwarding trees. We create forwarding trees using group tables~\cite{openflow-1.1.0}. {\name} uses a RESTful API to communicate with the end-points allowing them to specify their transfer properties and requirements (i.e., objective vectors) using which it computes and installs the required rules in the forwarding plane. We believe our techniques are easily applicable in today's inter-datacenter networks~\cite{b4, swan-backbone, facebook-express-backbone}. Our contributions can be summarized as follows:

\begin{itemize}
    \setlength{\itemsep}{0.5em}
    \item We propose a partitioning approach that reduces the effect of slow receivers by isolating them and attaching them using independent paths. We discuss various scenarios where this approach offers different levels of performance compared to the optimal partitioning on relaxed network topology and given max-min fair rate allocation.
    
    \item We incorporate binary objective vectors which allow applications to indicate transfer-specific objectives for receivers' completion times. Using the application-provided objective vectors, we can optimize for mixed completion time objectives based on the trade-off between total network capacity consumption and the receivers' average completion times.
    
    \item We present the {\name} heuristic, which computes a partitioning of receivers for every transfer given a binary objective vector. {\name} aims to minimize the completion time of receivers whose rank is indicated by applications/users with a one in the objective vector while saving as much bandwidth as possible by grouping receivers whose ranks are indicated with consecutive zeros in the objective vector.
    
    \item We perform extensive simulations and Mininet emulations with {\name} using synthetic and real-world Facebook inter-datacenter traffic patterns over large WAN topologies. Simulation results show that {\name} speeds up transfers to a small number of receivers~(e.g., $\geq 8$ receivers) by $\ge 2\times$ on the average completion time while the bandwidth used is $\leq 1.13\times$ compared to state-of-the-art. Transfers with more receivers receive larger benefits. For transfers to at least $16$ receivers, $75\%$ of the receivers complete at least $5\times$ faster and the fastest receiver completes $2.5\times$ faster compared to state-of-the-art. Compared to performing multicast as multiple unicast transfers with shortest path routing, {\name} reduces mean completion times by about $2\times$ while using $0.66\times$ of the bandwidth. Finally, Mininet emulations show that {\name} reduces the maximum group table entries needed by up to $3\times$.
\end{itemize}

\section{Background and related Work}

\noindent\textbf{Point to Multipoint (P2MP) Bulk Transfers:}
Similar to bulk multicast, P2MP transfers push data and content from one location to multiple locations. The transfer size, the source and the receivers are known and fixed prior to initiation of the transfer. Load-aware forwarding trees~\cite{dccast,quickcast,noormohammadpour2018fast,dartree}, flexible source selection \cite{AGE}, and store-and-forward~\cite{dtb, netstitcher} have been used in recent work to save bandwidth and speed up transfer completion. Several works around this subject focus on admission control for multicast transfers with deadlines \cite{AGE, dartree} which is orthogonal to our work in this paper. Other works focus on transferring large volumes of data with minimum increase in the network bandwidth cost \cite{dtb, netstitcher}. We build our work in this paper on top of recent work on optimizing the completion times of P2MP transfers \cite{dccast, quickcast}.

\vspace{0.5em}
\noindent\textbf{Network-layer Multicast:} 
A vast variety of network-layer multicast solutions have been proposed \cite{ip_multicast, centralized-multicast, tcp-smo, norm,mc_mc_bulk,rate_controlled_multicast,always_on_multicast}. In general, these solutions consider the dynamic scenarios where receivers may join or leave at any time; hence, they greedily adapt the multicast distribution tree as the receiver set evolves. The problem considered in this paper differs in the following key ways: we assume a known and fixed transfer size and set of receivers, we assume SDN-style visibility and control on the network routes, and we support for general completion time objectives. Compared to general multicast solutions that build multicast trees incrementally and greedily as new receivers join, given a known and fixed set of receivers, we can use the global knowledge of network load distribution and topology to select further optimized multicast trees that reduce the number of bottlenecks and improve the receiver completion times.



\vspace{0.5em}
\noindent\textbf{End-system Multicast:} 
These works form multicast trees in the application layer among the participating end-points \cite{rdcm, nice, AMMO, dc2, split-stream, overlay_hkust}; they have limited visibility and control of the underlying network, i.e., they cannot easily change routes, identify available bandwidth along network paths, etc. Therefore, depending on the network topology and the underlying routing approach, these techniques may offer solutions far from the optimal in minimizing the completion times of receivers.

\vspace{0.5em}
\noindent\textbf{Datacenter Multicast:} 
Recently, some works propose using multicast trees within and between datacenters \cite{datacast, avalanche, mctcp}. These approaches, however, operate specifically on datacenter networks that have regular and structured topologies whereas inter-datacenter networks are usually neither structured nor regular. Besides, these solutions do not aim at optimizing the completion times of multicast receivers.

\vspace{0.5em}
\noindent\textbf{Reliable transport protocols for multicast:}
Reliable multicast schemes ensure that data is completely and correctly received by all receivers using a variety of techniques such as FEC codes \cite{fec, alc, norm, avalanche_code, raptor, tornado}, retransmissions \cite{tcp-xm, mctcp, norm}, etc. Lost data can be detected using ACKs or NACKs. {\name} is orthogonal to and can use any multicast transport protocol.

\vspace{0.5em}
\noindent\textbf{Bit Indexed Explicit Replication:} 
A recent proposal, BIER~\cite{bier}, encodes forwarding state in the packet headers which simplifies network forwarding and improves scalability on large networks. BIER allows changes to multicast trees with a small overhead. {\name} can adopt BIER for creating and updating multicast trees to reduce the cost of rule installations and updates from SDN Group Tables.

\vspace{0.5em}
\noindent\textbf{Store-and-Forward (SnF):} 
SnF techniques~\cite{netstitcher, mbdt, dtb, mbdt_initial} have been proposed for unicast delay tolerant transfers to avoid periods of network congestion; a recent work called BDS uses SnF for bulk multicast over unicast TCP connections that connect receivers in a line~\cite{overlay_hkust} to increase network utilization given diurnal patterns of available capacity on backbone links. SnF, however, increases the protocol complexity and can incur additional bandwidth and storage costs on intermediate datacenters. Besides, formulating the bulk multicast problem using SnF will result in a completely new model as the data transmission rate across the edges of a multicast tree may be different which can change the nature of the problem in several ways. First, additional optimization metrics should be considered, among which are the total network storage and maximum per-node storage budget over multicast trees. Next, in case there is no limit on how much data can be stored per-node in a multicast tree, the slow receivers will automatically be isolated from the fast receivers over the tree. However, given that we most likely will have storage budgets per intermediate node on a multicast tree, this may create a complex relationship between the download speeds of fast and slow receivers. Therefore, modeling bulk multicasting with SnF can generate highly complex linear programs solving which may be slow. Application of SnF for bulk multicasting to optimize receiver completion times is considered part of the future work.

\vspace{0.5em}
\noindent\textbf{Peer-to-Peer File Distribution:} 
These techniques \cite{promise, bittorrent, slurpie, push_to_peer} function locally and greedily and cannot take advantage of SDN-style visibility and control of the inter-datacenter networks for global optimization of transfers among datacenters.

\section{System Model} \label{system_model}
We focus on proprietary inter-datacenter networks that connect geographically dispersed datacenters such as Microsoft Global WAN \cite{swan-backbone}, Facebook Express Backbone \cite{facebook-express-backbone} and Google GScale \cite{b4}. These networks are managed by one organization and their forwarding state can be managed in a logically centralized manner. A traffic engineering server that runs {\name} algorithm decides how traffic is forwarded in-network similar to other related work \cite{b4, swan, tempus, owan}. 

Cross datacenter traffic can, in general, be categorized as high priority user traffic that is highly sensitive to latency and internal traffic (also known as elastic or background traffic \cite{swan, tempus}) that is more resilient to latency. Internal traffic constitutes the majority of cross datacenter traffic, consists of huge volumes of replicated data and content that generate long-running transfers, and is growing at a much faster pace than user-generated traffic \cite{facebook-express-backbone}. By forwarding such traffic according to transfer properties (i.e., end-points and volume) and network topology (i.e., connectivity and available bandwidth) we can optimize some network-wide utility. Wide-area traffic management is a complex problem and in general a variety of metrics can be considered for optimization \cite{wide_area_tm}. We focus on internal traffic that is a result of data and content replication which can be modeled as bulk multicast transfers. These transfers are processed in an online manner as they arrive with the main objective of optimizing completion times. Also, forwarding entries, which are installed for every transfer upon arrival, are fixed until transfers' completion and are only updated in case of failures. Finally, we assume that all multicast transfers are of the same priority. Extension of the proposed solutions to a scenario where transfers are of different value to the operator/client(s) is considered as part of the future work.

We consider max-min fair \cite{max-min-fairness} rate allocation across multicast forwarding trees. Traffic is transmitted with the same rate from the source to all the receivers attached to a forwarding tree. To reach max-min fair rates, such rates can either be computed centrally over specific time periods, i.e., timeslots, and then be used for end-point traffic shaping or end-points can gradually converge to such rates in a distributed fashion in a way similar to TCP \cite{mctcp} (fairness is considered across trees). In our evaluations, we will consider the former approach for increased network utilization. Using a fair sharing policy addresses the starvation problem (such as in SRPT policy) and prevents larger transfers from blocking edges (such as in FCFS policy). Recent work has also shown that such conditions can worsen over trees \cite{quickcast}.

We use the notion of \textbf{objective vectors} to allow applications to define transfer-specific requirements which in general can improve overall system performance and reduce bandwidth consumption. An objective vector for a transfer is a vector of zeros and ones which is the same size as the number of receivers of that transfer. From left to right, the binary digit $i$ in this vector is associated with the $i\textsuperscript{th}$ fastest receiver. A one in the objective vector indicates that we are specifically interested in the completion time of the receiver associated with that rank in the vector. By assigning zeros and ones to different receiver ranks, it is possible to respect different applications' preferences or requirements while allowing the system to optimize bandwidth consumption further. The application/user, however, needs not be aware of the mapping between the downlink speeds (rank in the objective vector) and the receivers themselves.

Table \ref{table_omega} offers several examples. For instance, an objective vector of $\{0,0,0,1,0,0,0,0\}$ indicates the application's interest in the fourth fastest receiver. To respect the application's objective, we initially isolate the fourth receiver and do not group it with any other receiver. The first three fastest receivers can be grouped into a partition to save bandwidth. The same goes for the four slowest receivers. However, we do not group all receivers indicated with zeros into one partition initially (i.e., the top three receivers and the bottom four) to avoid slowing some of them down unnecessarily (in this case, the top three receivers). This forms the basis for the partitioning technique proposed in \S \ref{partitioning} that operates by building a hierarchy with multiple layers, where each layer is a valid partitioning solution, and selects the layer that gives the smallest average receiver completion times.

Although the objective vector can, in general, be any binary string of zeros and ones, it is worth noting that not all such combinations lead to meaningful objectives for datacenter applications. For example, an objective vector of $\{0,1,0,1,0,1,0,1\}$ may be unlikely to be used by an application. Having the ability to define and enforce any objective function though makes the system highly configurable and adaptable. Operators may come up with a set of rules based on which they can decide whether the objective vector proposed by an application is meaningful, or propose changes to a submission that is not deemed useful.

\begin{table}[t]
\caption{Behavior of Several Objective Vectors} \label{table_omega}
\centering
\footnotesize
\begin{tabular}{ |p{4cm}|p{8cm}| }
    \hline
    \textbf{Objective Vector ($\omega$)} & \textbf{Outcome (given $n$ receivers)} \\
    \hline
    \hline
    \raisebox{-0.6\totalheight}{\includegraphics[height=2.5em]{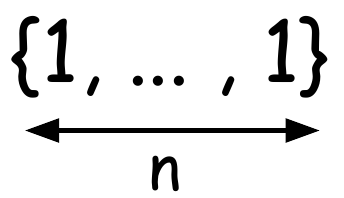}} & Interested in completion times of all individual receivers \\
    \hline
    \raisebox{-0.9\totalheight}{\includegraphics[height=2.5em]{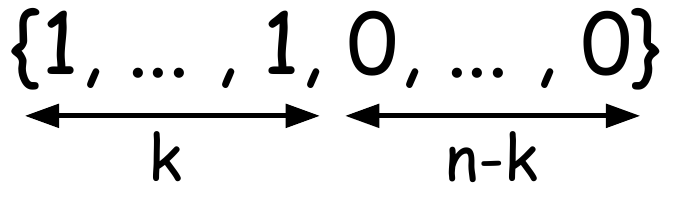}} & Interested in completion times of the top $k$ receivers (groups the bottom $n-k$ receivers to save bandwidth) \\
    \hline
    \raisebox{-1.2\totalheight}{\includegraphics[height=2.5em]{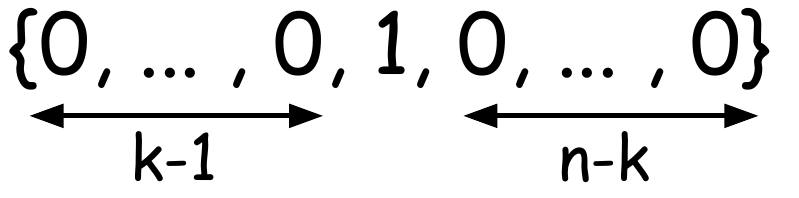}} & Interested in the completion time of the $k$\textsuperscript{th} receiver (groups the top $k-1$ receivers into a fast partition, and the bottom $n-k$ receivers into a slow one to save bandwidth) \\
    \hline
    \raisebox{-0.9\totalheight}{\includegraphics[height=2.5em]{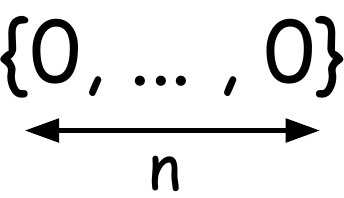}} & Not interested in the completion time of any specific receiver (all receivers form a single partition) \\
    \hline
\end{tabular}
\end{table}

\vspace{0.5em}
\noindent\textbf{Problem Statement:}
Given an inter-datacenter topology with known available bandwidth per link, the traffic engineering server is responsible for \textit{partitioning receivers} and \textit{selecting a forwarding tree per partition} for every incoming bulk multicast transfer. A bulk multicast transfer is specified by its source, set of receivers and volume of data to be delivered. The primary objective is minimizing average receiver completion times. In case an objective vector is specified, we want to minimize average completion times of receivers whose ranks are indicated with a $1$ in the vector as well as receivers indicated with consecutive $0$'s in the vector together as groups (receivers noted with consecutive $0$'s use the same forwarding tree and will have the same completion times). Minimizing bandwidth consumption, which is directly proportional to the size of selected forwarding trees, is considered a secondary objective.

\subsection{Online Greedy Optimization Model} \label{greedy_online}
The online bulk multicast partitioning and forwarding tree selection problem can be formulated using Eq. \ref{opt}-\ref{const_cap} with the added constraint that our rate allocation is max-min fair across forwarding trees for any selection of the partitions and the trees. Table \ref{table_var} lists the variables used in the formulation below.

\begin{table}
\caption{Definition of Variables} \label{table_var}
\centering
\footnotesize
\begin{tabular}{ |p{3cm}|p{10cm}| }
    \hline
    \textbf{Variable} & \textbf{Definition} \\
    \hline
    \hline
    $t_{now}$ & Current timeslot \\
    \hline
    $e$ & A directed edge \\
    \hline
    $C_e$ & Capacity of $e$ in bytes/second \\
    \hline
    $B_e(t)$ & Available bandwidth on edge $e$ at timeslot $t$ after setting aside usage of high priority user traffic \\
    \hline
    $B_e$ & Average available bandwidth on edge $e$ \\
    \hline
    $G$ & The directed inter-datacenter graph \\
    \hline
    $T$ & A directed Steiner tree \\
    \hline
    $\bm{\mathrm{V}}_G$ and $\bm{\mathrm{V}}_T$ & Set$\langle\rangle$ of vertices of $G$ and $T$ \\
    \hline
    $\bm{\mathrm{E}}_G$ and $\bm{\mathrm{E}}_T$ & Set$\langle\rangle$ of edges of $G$ and $T$ \\
    \hline
    $r_T(t)$ & The transmission rate over tree $T$ at $t$ \\
    \hline
    $\delta$ & Duration of a timeslot \\
    \hline
    $R$ & A bulk multicast transfer request \\
    \hline
    $S_{R}$ & Source datacenter of request $R$ \\
    \hline
    $A_{R}$ & Arrival time of request $R$ \\
    \hline
    $\mathcal{V}_{R}$ & Original volume of request $R$ \\
    \hline
    $\bm{\mathrm{D}}_{R}$ & Set$\langle\rangle$ of destinations of request $R$ \\
    \hline
    $\bm{\mathrm{R}}$ & Set$\langle\rangle$ of ongoing transfers \\
    \hline
    $P$ & A receiver partition of some request \\
    \hline
    $\bm{\mathrm{P}}_{R}$ & Set$\langle\rangle$ of partitions of some request $R$ \\
    \hline
    $T_{P}$ & The forwarding tree of partition $P$ \\
    \hline
    $\mathcal{V}_{P}^{[res]}$ & Current residual volume of partition $P$ of request $R$ \\
    \hline
    $\kappa_P$ & Estimated minimum completion time of partition $P$ \\
    \hline
    $L_e$ & Edge $e$'s total outstanding load (see \S \ref{forwardingtree}) \\
    \hline
    $\omega_R$ & Objective vector assigned to request $R$ \\
    \hline
    $\omega_R^{\star}$ & Weighted completion time vector computed from $\omega_R$ by replacing the last zero in a pack of consecutive zeros with the number of consecutive zeros in that pack (e.g., $\omega_R=\{0,0,0,1,0,0\} \rightarrow \omega_R^{\star}=\{0,0,3,1,0,2\}$) \\
    \hline
    $t_{\bm{\mathrm{D}}_{R}}$ & Vector of completion times of receivers of request $R$ sorted from fastest to slowest \\
    \hline
\end{tabular}
\end{table}

The set $\bm{\mathrm{R}}$ includes both the new transfer $R_N$ and all the ones already in the system for which we already have the partitions and forwarding trees. The optimization objective of Eq. \ref{opt} is to minimize the weighted sum of completion times of receivers of all requests $R \in \bm{\mathrm{R}}$ according to their objective vectors, and the total bandwidth consumption of $R_N$ by partitioning its receivers and selecting their forwarding trees (indicated by the term $\sum_{P \in \bm{\mathrm{P}}_{R_N}} \mathcal{V}_{P} \vert T_{P} \vert$). Operators can choose the non-negative coefficient $\epsilon$ according to the overall system objective to give a higher weight to minimizing the weighted completion time of receivers than reducing bandwidth consumption. Eq. \ref{const_dem} shows the demand constraints which state that the total sum of transmission rates over every tree for future timeslots is equal to the remaining volume of data per partition (each partition uses one tree). Eq. \ref{const_cap} presents the capacity constraints which state that the total sum of transmission rates per timeslot for all trees that share a common edge has to not go beyond its available bandwidth.

\setlength{\FrameSep}{2pt}
\begin{framed}
{\small
\begin{align}
    \textrm{\textbf{min}}~~~&\sum_{R \in \bm{\mathrm{R}}} \Big( t_{\bm{\mathrm{D}}_{R}} \cdot \omega_R^{\star} \Big) + \epsilon \sum_{P \in \bm{\mathrm{P}}_{R_N}} \mathcal{V}_{P} \vert T_{P} \vert \label{opt}\\
    \textrm{\textbf{S.t.}}~~~&\sum_{t} r_{T_P}(t) = \mathcal{V}_{P}^{[res]} \qquad \qquad \quad \forall P \in \bm{\mathrm{P}}_{R}, R \in \bm{\mathrm{R}} \label{const_dem} \\
    &\sum_{\{P \vert e \in T_P\}} r_{T_P}(t) \le B_e(t) \qquad \forall t, e, P \in \bm{\mathrm{P}}_{R}, R \in \bm{\mathrm{R}} \label{const_cap}
\end{align}}
\end{framed}

This online discrete optimization problem is highly complex as it is unclear how receivers should be partitioned into multiple subsets to reduce completion times and there is an exponential number of possibilities. Selection of forwarding trees to minimize completion times is also a hard problem. In \S \ref{Iris}, we will present a heuristic that approximates a solution to this optimization problem inspired by the findings in \S \ref{partitioning_model}.

\section{Partitioning of Receivers on a Relaxed Topology} \label{partitioning_model}
\begin{figure}
    \centering
    \includegraphics[width=0.5\columnwidth]{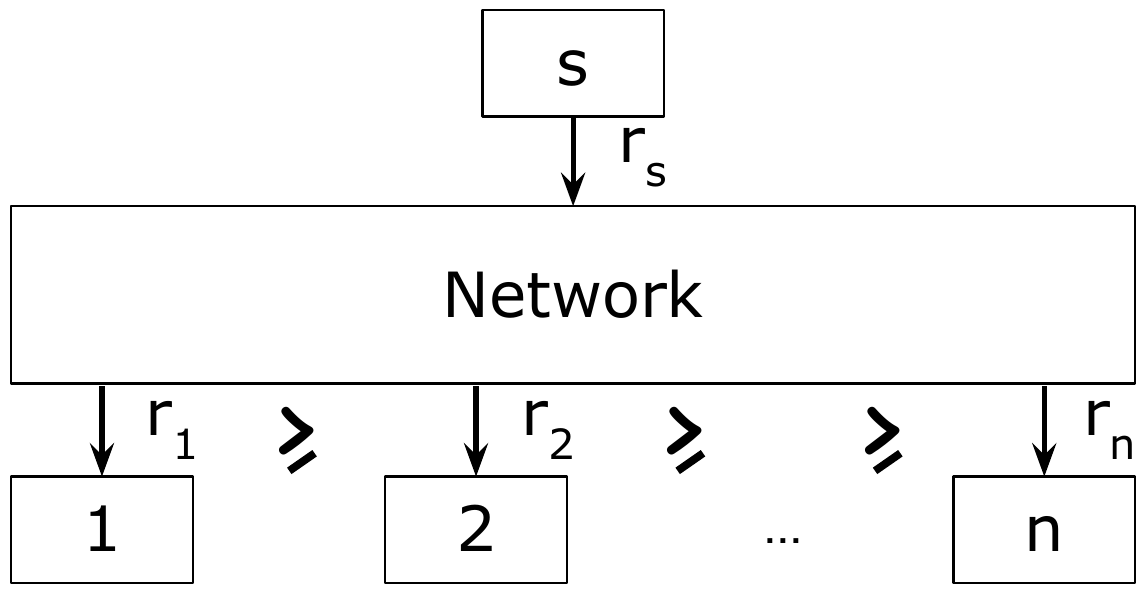}
    \caption{A relaxed topology with infinite core capacity, and uplink and downlink capacities of $r_s$ and $r_1 \ge \dots \ge r_n$.}
    \label{fig:problem_formulation}
\end{figure}

Due to the high complexity of the partitioning problem as a result of physical topology, we first study a relaxed topology where every datacenter is attached with a single uplink/downlink to a network with infinite core capacity and so the network core cannot become a bottleneck. As shown in Figure \ref{fig:problem_formulation}, the sender has a maximum uplink rate of $r_s$ and transmits to a set of $n$ receivers with different maximum downlink rates of $r_i, \forall i \in \{1,\dots,n\}$. In \S \ref{forwardingtree}, we discuss a load-balancing forwarding tree selection approach that aims to distribute load across the network to minimize the effect of bottlenecks within the network core.

Without loss of generality, let us also assume that the receivers in Figure \ref{fig:problem_formulation} are sorted by their downlink rates in descending order. The sender can initiate multicast flows to any partition, i.e., a subset of receivers, given that every receiver appears in exactly one partition. All receivers in a partition will have the same multicast rate that is the rate of the slowest receiver in the partition. To compute rates at the uplink, we consider the max-min fair rate allocation policy as stated earlier in \S \ref{system_model}. In this context, we would like to compute the number of partitions as well as the receivers that should be grouped per partition to minimize mean completion times.

\vspace{0.5em}
\textbf{Theorem 1.} \textit{Given receivers sorted by their downlink rates, partitioning that groups consecutive receivers is pareto-optimal with regards to minimizing completion times.}

\vspace{0.5em}
\textbf{\textit{Proof.}} We use proof by contradiction. Let us assume a partitioning where non-consecutive receivers are grouped together, that is, there exist two partitions $P_1$ and $P_2$ where part of partition $P_1$ falls in between receivers of $P_2$ or the other way around. Let us call the slowest receivers of $P_1$ and $P_2$ as $j_1$ and $j_2$, respectively. Across $j_1$ and $j_2$, let us pick the fastest and call it $f(j_1,j_2)$. If $f(j_1,j_2) = j_1$ (i.e., in the non-decreasing order of downlink speed from left to right, $P_2$ appears before $P_1$ as in $P_2\{\dots\}~P_1\{\dots,j_1\}~P_2\{\dots,j_2\}~\dots$), then by swapping the fastest receiver in $P_2$ and $j_1$, we can improve the rate of $P_1$ while keeping the rate of $P_2$ the same. If $f(j_1,j_2) = j_2$, then by swapping the fastest receiver in $P_1$ and $j_2$, we can improve the rate of $P_2$ while keeping the rate of $P_1$ the same. This can be done in both cases without changing the number of partitions, or number of receivers per partition across all partitions. Since the new partitioning has a higher or equal achievable rate for one of the partitions, the total average completion times will be less than or equal to that of original partitioning, which means the original partitioning could not have been optimal.

\subsection{Our Partitioning Approach}
\begin{figure}
    \centering
    \includegraphics[width=\columnwidth]{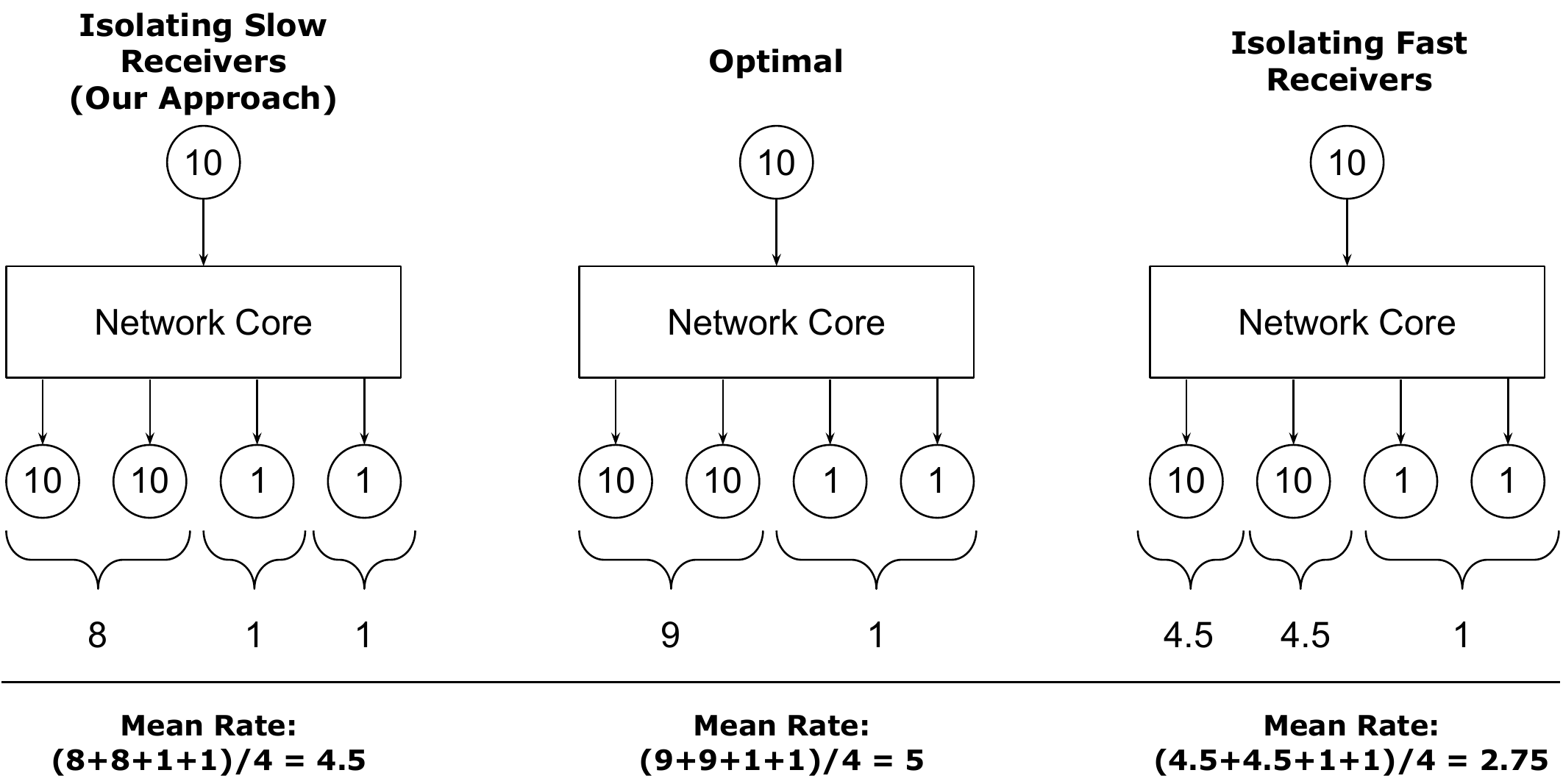}
    \caption{Various partitioning solutions for a scenario with four receivers. Numbers show the downlink and uplink speeds of nodes and curly brackets indicate the partitions where all nodes in a partition receive data at the same rate. The objective is to maximize the average rate of receivers given the max-min fairness policy.}
    \label{fig:partitioning_example_1}
\end{figure}

Based on Theorem 1, the number of possible partitioning scenarios that can be considered for minimum average completion times is the number of compositions of integer $n$, that is, $2^{n-1}$ ways which can be a large space to search. To reduce complexity, we propose to isolate slow receivers from the rest of receivers to minimize their effect. In other words, given an integer $1 \le M \le n$, we propose to group the first $n - M + 1$ fastest receivers into one partition and the rest of the receivers as separate $1$-receiver partitions ($M - 1$ in total). Since we do not know the value of integer $M$, we will try all possible values, that is, $n$ in total which will help us find the right threshold for the separation of fast and slow receivers. In particular, we compute the total average downlink rate of all receivers for the given transfer for every value of $M$ and select the $M$ that maximizes the average rate.\footnote{Or alternatively minimizes the average completion times of receivers.} As shown in Figure \ref{fig:problem_formulation}, the uplink at the sender has a rate of $r_s$ which will be divided across all the multicast flows that deliver data to the receivers. Isolating a slow receiver only takes a small fraction of the sender's uplink which is why this technique is effective as we will later see in evaluations. An example of this approach and how it compares with the optimal solution is shown in Figure \ref{fig:partitioning_example_1} where our solution selects $M=3$ partitions isolating the two slow receivers.

\begin{figure}
    \centering
    \includegraphics[width=0.8\columnwidth]{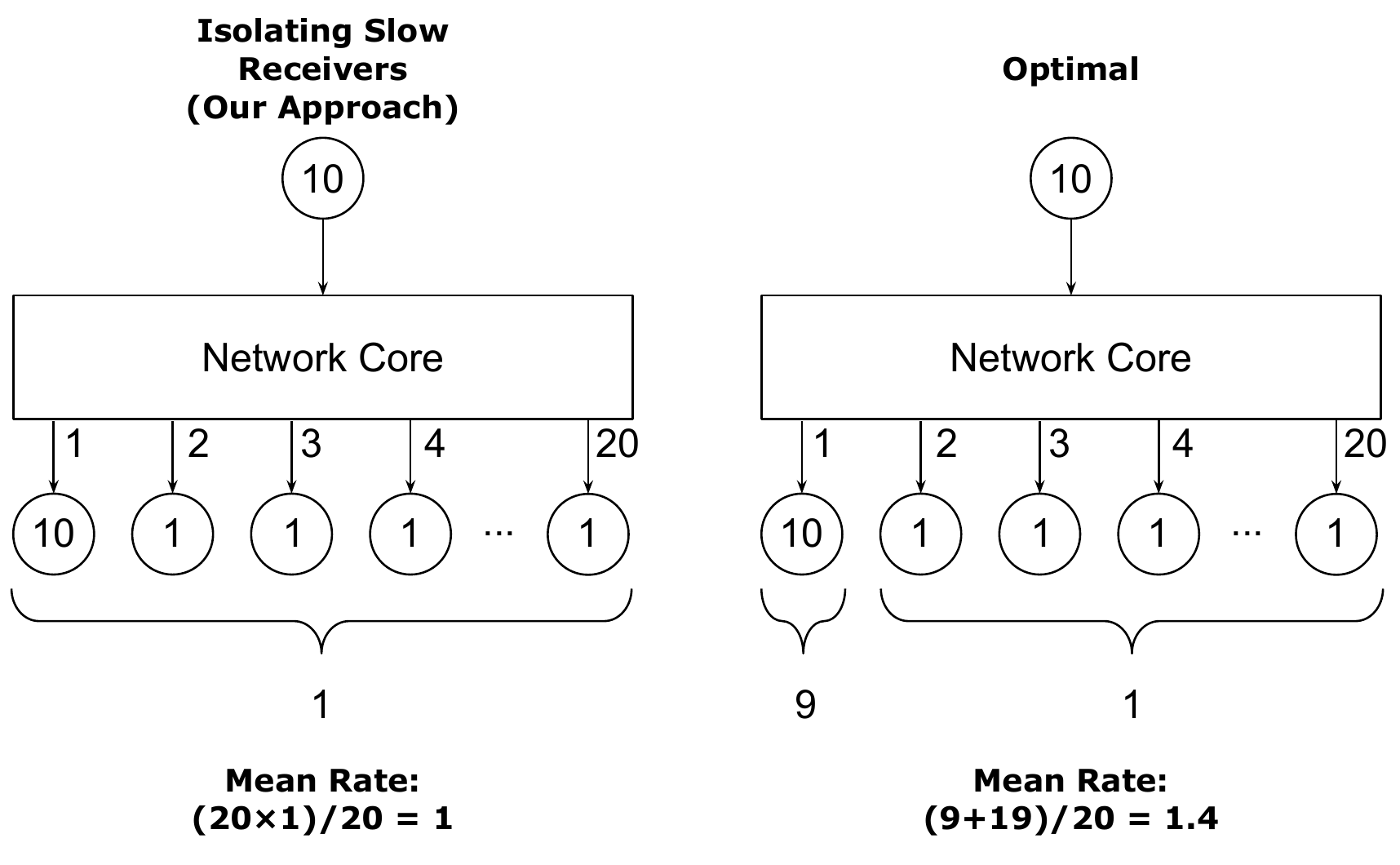}
    \caption{A worst-case scenario for the proposed partitioning scenario. Numbers within the nodes show the downlink and uplink speeds of nodes and curly brackets indicate the partitions where all nodes in a partition receive data at the same rate. The objective is to maximize the average rate of receivers given the max-min fairness policy.}
    \label{fig:partitioning_example_2}
\end{figure}

A main determining factor in the effectiveness of this approach is how $r_s$ compares with $\sum_{1 \le i \le n} r_i$. If $r_s$ is larger, then simply using $n$ partitions will offer the maximum total rate to the receivers. The opposite is when $r_s \ll \sum_{1 \le i \le n} r_i$ in which case using a single partition offers the highest total rate. In other cases, given the partitioning approach mentioned above, the worst-case scenario happens when there are many slow receivers and only a handful of fast receivers. An example has been shown in Figure \ref{fig:partitioning_example_2}. In the scenario on the left, our approach groups all the receivers into one partition where they all receive data at the rate of one. That is because by isolating slow receivers we can either get a rate of one or less than one if we isolate more than nine slow receivers, which means using one partition is enough. The optimal case, however, groups all the slow receivers into one partition. In general, scenarios like this rarely happen as the number of slow receivers over inter-datacenter networks is usually small, i.e., most datacenters are connected using high capacity links with large available bandwidth.\footnote{We have deduced this by looking at many WAN topologies available on the Internet Topology Zoo \cite{topologyzoo}. We found that in most topologies, a small fraction of nodes are connected using significantly slower links while the variation of downlink/uplink capacity for the rest of the nodes is not significant.} In general, since we consider all values of $M$ from $1$ to $n$ partitions, the solution obtained from our partitioning approach cannot be worse than the two baseline approaches of using a single multicast tree for all receivers and unicasting to all receivers using separate paths.

\subsection{Incorporating Objective Vectors}
\begin{figure}
    \centering
    \includegraphics[width=0.8\columnwidth]{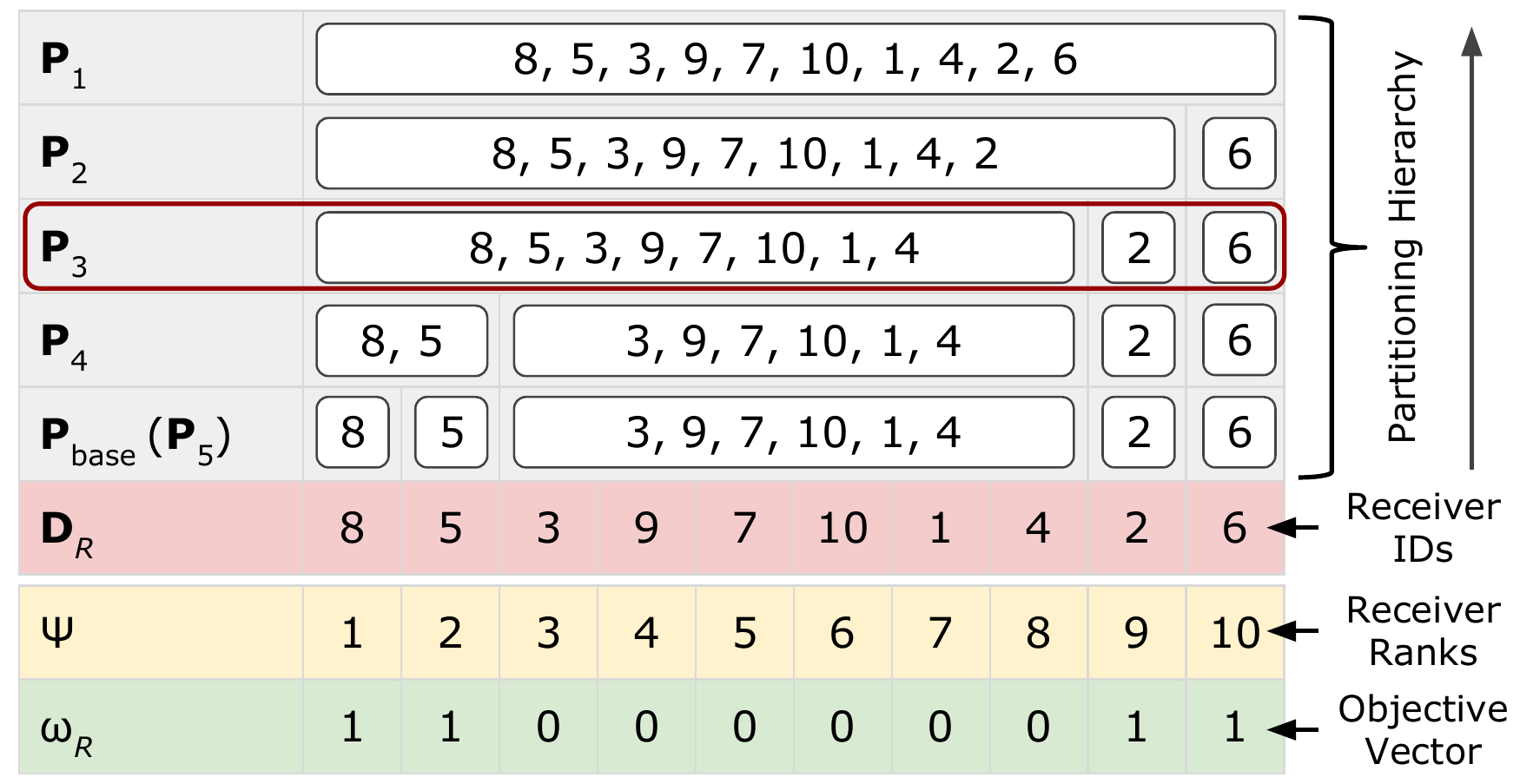}
    \caption{Example of a partitioning hierarchy for a transfer with 10 receivers (the topology not shown).}
    \label{fig:wancast_clustering_example}
\end{figure}

We allow users to supply an objective vector along with their multicast transfers to better optimize the network performance, that is, total network capacity consumption and receiver completion times. We incorporate the objective vectors by grouping receivers with consecutive ranks that are indicated with zeros in the objective vector and treating them as one partition in the whole process. That is because the users have indicated no interest in the completion times of those receivers, so we might as well reduce the network capacity usage by grouping them from the beginning. Figure \ref{fig:wancast_clustering_example} shows an example of building possible solutions by isolating slow receivers and incorporating the user-supplied objective vector, which we refer to as the partitioning hierarchy. Please note that this hierarchy moves in the reverse direction, that is, instead of isolating slow receivers, it merges fast receivers from bottom to the top.

Each layer in this hierarchy, labeled as $\bm{\mathrm{P}}_{i}, 1 \le i \le 5$, represents a valid partitioning solution.\footnote{The associated network topology is not shown.} We see that receivers indicated with consecutive zeros in $\omega_{R}$ are merged into one big partition at the base layer or $\bm{\mathrm{P}}_{5}$. Also, we see that as we move up, the two fastest partitions at each layer are merged, which reduces total bandwidth consumption. For each layer, we compute the average completion time of receivers and then select the layer that offers the least value, in this case, $\bm{\mathrm{P}}_{3}$ was chosen.

\section{Iris} \label{Iris}
We apply the partitioning technique discussed in the previous chapter to real-world inter-datacenter networks. We develop a heuristic for partitioning receivers on real-world topologies without relaxations of \S \ref{partitioning_model}. We will generate multiple valid partitioning solutions in the form of a hierarchy where layers of the hierarchy present feasible partitioning solutions and each layer is formed by merging the two fastest partitions of the layer below.\footnote{In general, it is not possible to offer optimality guarantees due to the highly varying factors of network topology, transfer arrivals, and the distribution of transfer volumes. However, our extensive simulations in \S \ref{evaluations} show that our approach can offer significant improvement on other approaches over various topologies and traffic patterns. Also, as a result of building a hierarchy of partitioning options and selecting the best one, our solution will be at least as good as either using a single multicast tree or using unicasting to all receivers.}

\begin{figure}
    \centering
    \includegraphics[width=\columnwidth]{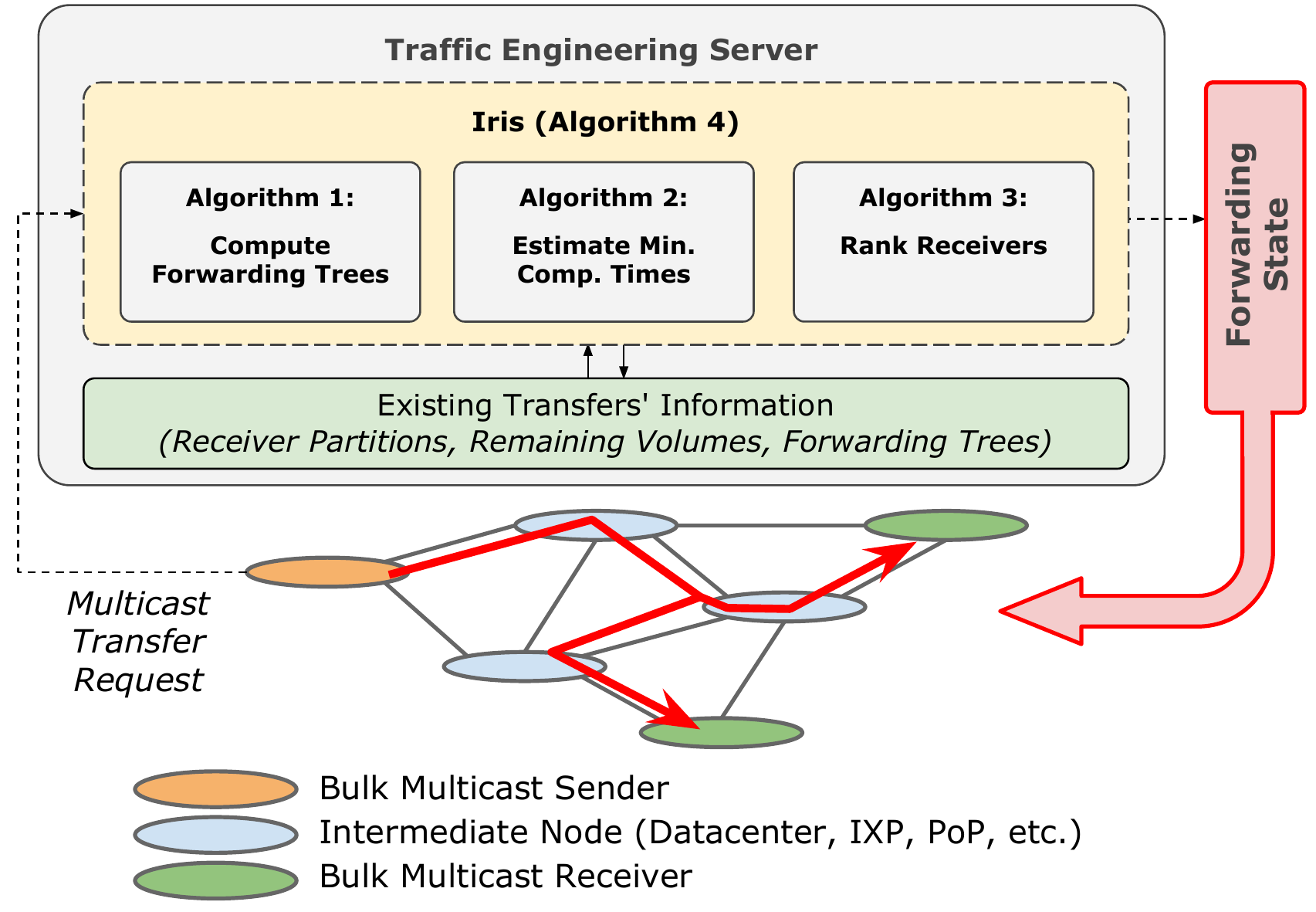}
    \caption{Pipeline of {\name}.}
    \label{fig:iris_pipeline}
\end{figure}

We present {\name}, a heuristic that runs on the traffic engineering server to manage bulk multicast transfers.\footnote{Unicast transfers are a special case with a single receiver.} When a bulk multicast transfer arrives at an end-point, it will communicate the request to the traffic engineering server which will then invoke {\name}. It uses the knowledge of physical layer topology, available bandwidth on edges after deducting the share of high priority user traffic and other running transfers to compute partitions and forwarding trees. The traffic engineering server pulls end-points' actual progress periodically to determine their exact remaining volume across transfers to compute the total outstanding load per edge for all edges. {\name} consists of four modules as shown in Figure \ref{fig:iris_pipeline} which we discuss in the following subsections. {\name} aims to find an approximate solution to the optimization problem of Eq. \ref{opt} assuming $\epsilon \ll 1$ to prioritize minimizing completion times over minimizing bandwidth consumption. We will empirically evaluate {\name} by comparing it to recent work and a lower bound in \S \ref{evaluations}.

\subsection{Choosing Forwarding Trees} \label{forwardingtree}
Load aware forwarding trees are selected given the link capacity information on the topology and according to other ongoing bulk multicast transfers across the network to reduce the completion times by mitigating the effect of bottlenecks. Tree selection should also aim to keep bandwidth consumption low by minimizing the number of edges per tree where an edge could refer to any of the links on the physical topology. To select a forwarding tree, a general approach that can capture a wide range of selection policies is to assign weights to edges of the inter-datacenter graph $G$ and select a minimum weight Steiner tree \cite{steiner_tree_problem}. Per edge $e \in \bm{\mathrm{E}}_G$, we assume a virtual queue that increases by volume of every transfer scheduled on that edge and decreases as traffic flows through it. Since edges differ in capacity, completing the same virtual queue size may need significantly different times for different links. We extend the metric used in a recent work \cite{quickcast} that is called load $L_{e}$ as follows.

\begin{equation}
    L_e = \frac{1}{B_e} \sum_{\{P \in \bm{\mathrm{P_{R}}} \vert e \in T_{P}\}} \mathcal{V}_{P}^{[res]} \label{l_e}
\end{equation}

In Eq. \ref{l_e}, $\bm{\mathrm{P_{R}}}$ is the set of partitions of receivers of all ongoing transfers. This equation sums up the remaining volumes of all trees that use a specific edge (total virtual queue size) and divides that by the average available bandwidth on that edge to compute the minimum possible time it takes for all ongoing transfers on that edge to complete. In the tree selection process, to keep completion times low, we need to avoid edges for which this value is large.

With this metric available, to select a forwarding tree given a sender and several receivers, we will first assign an edge weight of $W_e = L_e + \frac{\mathcal{V}_{R}}{B_e}$ to all edges and then select a minimum weight Steiner tree as shown in Algorithm \ref{tree_algorithm}. With this edge weight, compared to edge utilization which has been extensively used in literature for traffic engineering, we achieve a more stable measure of how busy a link is expected to be in the near future on average. We considered the second term in edge weight to reduce total bandwidth use when there are multiple trees with the same weight. It also leads to the selection of smaller trees for larger transfers which decreases the total bandwidth consumption of {\name} further in the long run.

\vspace{0.5em}
\noindent\textbf{Complexity:} To compute a minimum weight Steiner tree we use a heuristic that is called GreedyFLAC \cite{Watel2014} which given the set of terminal nodes $\Gamma$, has a guaranteed polynomial running time of $\mathcal{O}(\vert \bm{\mathrm{V}}_G \vert \vert \bm{\mathrm{E}}_G \vert + \vert \bm{\mathrm{V}}_G \vert^2 log(\vert \bm{\mathrm{V}}_G \vert) \vert \Gamma \vert + \vert \bm{\mathrm{V}}_G \vert^2 \vert \Gamma \vert^3)$.

\SetAlgoVlined
\SetInd{1.2em}{0.5em}
\begin{algorithm}[t]
\caption{Compute A Forwarding Tree} \label{tree_algorithm}
\SetKw{KwBy}{by}

\small

\KwIn{Steiner tree terminal nodes $\bm{\mathrm{\Gamma}}$, request $R$}

\KwOut{Edges of a tree}

\SetKwProg{CompForwardingTree}{CompForwardingTree}{}{}

\CompForwardingTree{$\mathrm{(}\bm{\mathrm{\Gamma}},R\mathrm{)}$}{
    
    \vspace{0.4em}
    To every edge $e \in \bm{\mathrm{E}}_G$, assign weight $W_e = (L_{e} + \frac{\mathcal{V}_{R}}{B_e})$\;
    
    \vspace{0.4em}
    \Return{A minimum weight Steiner tree that connects the nodes in set $\bm{\mathrm{\Gamma}}$ (we used a hueristic \cite{DSTAlgoEvaluation, Watel2014})}\;
}
\end{algorithm}

\subsection{Estimating Minimum Completion Times} \label{ct_estimation}
The purpose of this procedure is to estimate the minimum completion time of different partitions of a given transfer considering available bandwidth over the edges and applying max-min fair rate allocation when there are shared links across forwarding trees.  Algorithms \ref{rank_algorithm} and \ref{wancast_algorithm} then use the minimum completion time per partition to rank the receivers (i.e., faster receivers have an earlier completion time) and then decide which partitions to merge. Computing the minimum completion times is done by assuming that the new transfer request has access to all the available bandwidth and compared to computing the exact completion times is much faster. Besides, calculating the exact completion times is not particularly more effective due to the continuously changing state of the system as new transfer requests arrive. Since available bandwidth over future timeslots is not precisely known, we can use estimate values similar to other work \cite{netstitcher, tempus, amoeba}. Algorithm \ref{ct_calc_algorithm} shows how the minimum completion times are computed.

\vspace{0.5em}
\noindent\textbf{Complexity:} For a new request $R$ with $\vert \bm{\mathrm{P}} \vert$ partitions, this algorithm calls Algorithm \ref{tree_algorithm}, $\vert \bm{\mathrm{P}} \vert$ times. It then computes max-min fair rates per partition for timeslots until all partitions finish. Computing max-min fair rates per timeslot has a complexity of $\mathcal{O}(\vert \bm{\mathrm{P}} \vert \vert \bm{\mathrm{E}}_G \vert)$. This process continues for $\mathcal{O}(\frac{\vert \bm{\mathrm{P}} \vert~\mathcal{V}_R}{\min_{e,t} B_e(t)})$ iterations. Therefore, the complexity of this algorithm is $\mathcal{O}(\vert \bm{\mathrm{P}} \vert~( \mathcal{C}_{Algorithm~1} + \vert \bm{\mathrm{E}}_G \vert~\frac{\vert \bm{\mathrm{P}} \vert~\mathcal{V}_R}{\min_{e,t} B_e(t)}))$.

\SetAlgoVlined
\SetInd{1.2em}{0.5em}
\begin{algorithm}[t]
\caption{Minimum Completion Times} \label{ct_calc_algorithm}
\SetKw{KwBy}{by}

\small

\KwIn{A set of partitions $\bm{\mathrm{P}}$, request $R$}

\KwOut{Completion time of every partition in $\bm{\mathrm{P}}$}

\SetKwProg{MinimumCompletionTimes}{MinimumCompletionTimes}{}{}

\MinimumCompletionTimes{$\mathrm{(}\bm{\mathrm{P}},R\mathrm{)}$}{

    \vspace{0.4em}
    $\bm{\mathrm{f}} \gets \emptyset$, $t \gets t_{now}+1$\;
    
    \vspace{0.4em}
    $\gamma_P \gets \mathcal{V}_R,~\forall P \in \bm{\mathrm{P}}$\;
    
    \vspace{0.4em}
    $T_P \gets$ \texttt{CompForwardingTree}($P,R$), $\forall P \in \bm{\mathrm{P}}$\;
    
    \vspace{0.4em}
    \While{$\vert \bm{\mathrm{f}} \vert < \vert \mathrm{\bm{\mathrm{P}}} \vert$}{
        
        \vspace{0.4em}
        Compute $r_P(t),\forall P \in \{\bm{\mathrm{P}} - \bm{\mathrm{f}}\}$, max-min fair rate \cite{max-min-fairness} allocated to tree $T_P$ at timeslot $t$ given available bandwidth of $B_e(t)$ on every edge $e \in \bm{\mathrm{E}}_G$\;
        
        \vspace{0.4em}
        $\gamma_P \gets \gamma_P - \delta \times r_P(t),~\forall P \in \bm{\mathrm{P}}$\;
        
        \vspace{0.4em}
        \ForEach{$P \in \{\bm{\mathrm{P}} - \bm{\mathrm{f}}\}$}{
            \vspace{0.4em}
            \If{$\gamma_P = 0$}{
                \vspace{0.4em}
                $\kappa_P \gets t$, $\bm{\mathrm{f}} \gets \bm{\mathrm{f}} \cup P$\;
            }
        }
        
        \vspace{0.4em}
        $t \gets t+1$\;
    
    }
    
    \vspace{0.4em}
    \Return{$\kappa_P,\forall P \in \bm{\mathrm{P}}$}
}
\end{algorithm}

\subsection{Assigning Ranks to Receivers} \label{receiver_rank_assignment}
Algorithm \ref{rank_algorithm} assigns ranks to individual receivers according to their minimum completion times taking into account available bandwidth over edges as well as edges' load in the path selection process. This ranking is used along with the provided objective vector later to partition receivers.

\vspace{0.5em}
\noindent\textbf{Complexity:} This algorithm calls Algorithm \ref{ct_calc_algorithm} over all receivers as separate 1-node partitions, then sorts the nodes which gives a complexity of $\mathcal{O}(\mathcal{C}_{Algorithm~2} + \vert \bm{\mathrm{D}}_R \vert~log(\vert \bm{\mathrm{D}}_R \vert))$.

\SetAlgoVlined
\SetInd{1.2em}{0.5em}
\begin{algorithm}[t]
\caption{Assign Receiver Ranks} \label{rank_algorithm}
\SetKw{KwBy}{by}

\small

\KwIn{Request $R$}

\KwOut{$\psi_r$, i.e., rank of receiver $r \in \bm{\mathrm{D}}_R$}

\SetKwProg{AssignReceiverRanks}{AssignReceiverRanks}{}{}

\AssignReceiverRanks{$\mathrm{(}R\mathrm{)}$}{
    
    \vspace{0.4em}
    {\color{gray}/* Every receiver is treated as a separate partition */}
    
    \vspace{0.4em}
    $\{\kappa_r,~\forall r \in \bm{\mathrm{D}}_R\} \gets $ \texttt{MinimumCompletionTimes}($\bm{\mathrm{D}}_R, R$)\;
    
    \vspace{0.4em}
    $\psi_r \gets$ Position of receiver $r$ in the list of all receivers sorted by their estimated minimum completion times (fastest receiver is assigned a rank of $1$), $\forall r \in \bm{\mathrm{D}}_R$\;
    
    \vspace{0.4em}
    \Return{$\psi_r,\forall r \in \bm{\mathrm{D}}_R$}\;
}
\end{algorithm}

\SetAlgoVlined
\SetInd{1.2em}{0.5em}
\begin{algorithm}
\caption{Compute Receiver Partitions and Trees (\name)} \label{wancast_algorithm}
\SetKw{KwBy}{by}

\small

\KwIn{Request $R$, binary objective vector $\omega_R$}

\KwOut{Partitions of request $R$ and their forwarding trees}

\SetKwProg{CompPartitionsAndTrees}{CompPartitionsAndTrees}{}{}
\CompPartitionsAndTrees{$\mathrm{(}R,\omega_R\mathrm{)}$}{
    
    \vspace{0.4em}
    {\color{gray}/* Initial partitioning using the objective vector $\omega_R$ */}
    
    \vspace{0.4em}
    $\{\psi_r, \forall r \in \bm{\mathrm{D}}_R\} \gets$ \texttt{AssignReceiverRanks}($R$)\;
    
    \vspace{0.4em}
    $\bm{\mathrm{D}}_{R}^{s} \gets$ Receivers $r$ sorted by $\psi_r, \forall r \in \bm{\mathrm{D}}_R$ ascending\;
    
    \vspace{0.4em}
    $\bm{\mathrm{P}}_{base} \gets \{$Any receiver $r \in \bm{\mathrm{D}}_R$ for which $\omega_R<\psi_r>$ is $1$ as a separate partition$\} \cup \{$Group receivers that appear consecutively on $\bm{\mathrm{D}}_{R}^{s}$ for which $\omega_R<\psi_r>$ is $0$, each group forms a separate partition$\}$\;
    
    \vspace{0.4em}
    {\color{gray}/* Building the partitioning hierarchy for $\bm{\mathrm{P}}_{base}$ */}
    
    \vspace{0.4em}
    $\bm{\mathrm{P}}_{\vert \bm{\mathrm{P}}_{base} \vert} \gets$ $\bm{\mathrm{P}}_{base}$\;
    
    \vspace{0.4em}
    \For{$l = \vert \bm{\mathrm{P}}_{base} \vert$ \KwTo $l = 1$ \KwBy $-1$}{
        \vspace{0.4em}
        $\{\kappa_P,~\forall P \in \bm{\mathrm{P}}_l\} \gets $ \texttt{MinimumCompletionTimes}$\mathrm{(\bm{\mathrm{P}}_l,R)}$\;
        
        \vspace{0.4em}
        $\kappa_l \gets \frac{\sum_{P \in \bm{\mathrm{P}}_l} (\vert P \vert~\kappa_P)}{\vert \bm{\mathrm{D}}_R \vert}$; ~{\color{gray}/* Compute the best average completion times */}
        
        \vspace{0.4em}
        Assuming receivers are sorted from left to right by increasing order of rank, merge the two partitions on the left, $P$ and $Q$, to form $PQ$\;
        
        \vspace{0.4em}
        $\bm{\mathrm{P}}_{l-1} \gets \{PQ\} \cup \{\bm{\mathrm{P}}_{l}-\{P,Q\}\}$\;
    }
    
    \vspace{0.4em}
    Find $l_{min}$ for which $\kappa_{l_{min}} \le \min_{1 \le l \le \vert \bm{\mathrm{P}}_{base} \vert} \kappa_l$, if multiple layers have the same $\kappa_l$, choose the layer with minimum total weight over all of its forwarding trees, i.e., select $l_{min}$ to $\min \sum_{P \in \bm{\mathrm{P}}_{l_{min}}} (\sum_{e \in T_P} W_e)$\;
    
    \vspace{0.4em}
    {\color{gray}/* Compute forwarding trees for the partitions */}
    
    \vspace{0.4em}
    \ForEach{$P \in \bm{\mathrm{P}}_{l_{min}}$}{
        \vspace{0.4em}
        $T_{P} \gets$ \texttt{CompForwardingTree}($P,R$)\;
        
        \vspace{0.4em}
        \ForEach{$e \in T_{P}$}{
            \vspace{0.4em}
            $L_e \gets L_e + \frac{\mathcal{V}_R}{B_e}$, $W_e \gets W_e + \frac{\mathcal{V}_R}{B_e}$\;
        }
    }
    
    \vspace{0.4em}
    \Return{$(P,~T_P),~\forall P \in \bm{\mathrm{P}}_{l_{min}}$}\;
}
\end{algorithm}

\subsection{The \name~Algorithm} \label{partitioning}
The \name\ algorithm computes receiver partitions using hierarchical partitioning and assigns each partition a multicast forwarding tree. The partitioning problem is solved per transfer and determines the number of partitions and the receivers that are grouped per partition. {\name} uses a partitioning technique inspired by the findings of \S \ref{partitioning_model} that is computationally fast, significantly improves receiver completion times, and operates only relying on network topology and available bandwidth per edge.\footnote{The available bandwidth per edge is computed by deducting the quota for higher priority user traffic from the link capacity.} We refer to our approach as hierarchical partitioning as it builds a hierarchy of different partitioning solutions.

We build a partitioning hierarchy with numerous layers and examine the various number of partitions from bottom to the top of the hierarchy while looking at the average of minimum completion times. Given that each node in a real-world topology may have multiple interfaces, by building a hierarchy, we consider the discrete nature of forwarding tree selection on the physical network topology. The process consists of two steps as follows.

Algorithm \ref{wancast_algorithm} illustrates how {\name} partitions receivers with an objective vector. We first use the receiver ranks from Algorithm \ref{rank_algorithm} and the objective vector to create the base of partitioning hierarchy, $\bm{\mathrm{P}}_{base}$. We first sort the receivers by their ranks from fastest to slowest and then group them according to the weights in the objective vector. For any receiver whose rank in the objective vector has a value of $1$, we consider a separate partition (single node partition) which allows the receiver to complete as fast as possible by not attaching it to any other receiver. Next, we group receivers with consecutive ranks that are assigned a value of $0$ in the objective vector into partitions with potentially more than one receiver, which allows us to save as much bandwidth as possible since the user has not indicated interest in their completion times.

Now that we have a set of base partitions $\bm{\mathrm{P}}_{base}$, a heuristic creates a hierarchy of partitioning solutions with $\vert \bm{\mathrm{P}}_{base} \vert$ layers where every layer $1 \le l \le \vert \bm{\mathrm{P}}_{base} \vert$ is made up of a set of partitions $\bm{\mathrm{P}}_{l}$. Each layer is created by merging two partitions from the layer below going from the bottom to the top of hierarchy. At the bottom of the hierarchy, we have the base partitions. Also, at any layer, any partition $P$ is attached to the sender using a separate forwarding tree $T_{P}$. We first compute the average of minimum completion times of all receivers at the bottom of the hierarchy. We continue by merging the two partitions that hold receivers with highest ranks. When merging two partitions, the faster partition is slowed down to the speed of slower partition. A new forwarding tree is computed for the resulting partition using the forwarding tree selection heuristic of Algorithm \ref{tree_algorithm} to all receivers in that partition, and the average of minimum completion times for all receivers are recomputed. This process continues until we reach a single partition that holds all receivers. In the end, we select the layer at which the average of minimum completion times across all receivers is minimum, which gives us the number of partitions, the receivers that are grouped per partition, and their associated forwarding trees. If there are multiple layers with the minimum average completion times, the one with minimum total forwarding tree weight across its forwarding trees is chosen which on average leads to better load distribution.

\vspace{0.5em}
\noindent\textbf{Complexity:} This algorithm first calls Algorithm \ref{rank_algorithm}, then it calls Algorithm \ref{ct_calc_algorithm} up to $\vert \bm{\mathrm{D}}_R \vert$ times. At the end, it also runs Algorithm \ref{tree_algorithm} up to $\vert \bm{\mathrm{D}}_R \vert$ times. Therefore, this algorithm has a complexity of $\mathcal{O}(\mathcal{C}_{Algorithm~3} + \vert \bm{\mathrm{D}}_R \vert~(\mathcal{C}_{Algorithm~2} + \mathcal{C}_{Algorithm~1}))$.

\begin{table*}
\centering
{\footnotesize \caption{Various topologies and traffic patterns used in evaluation. One unit of traffic is equal to what can be transmitted at the rate of the fastest link over a given topology per timeslot.} \label{table_evaluations}
\begin{tabular}{p{2cm}|p{3cm}|p{8cm}|}
\cline{2-3}
 & \textbf{Name} & \textbf{Description} \\ \hline
\multicolumn{1}{|l|}{\multirow{2}{*}{Topology}} & GEANT & Backbone and transit network across Europe with 34 nodes and 52 links. Link capacity from 45 Mbps to 10 Gbps. \\ \cline{2-3} 
\multicolumn{1}{|l|}{} & UNINETT & Backbone network across Norway with 69 nodes and 98 links. Most links have a capacity of 1, 2.5 or 10 Gbps. \\ \hline
\multicolumn{1}{|l|}{\multirow{4}{*}{Traffic Pattern}} & Light-tailed & Based on Exponential distribution with a mean of $20$ units per transfer. \\ \cline{2-3} 
\multicolumn{1}{|l|}{} & Heavy-tailed & Based on Pareto distribution with the minimum of $2$ units, the mean of $20$ units, and the maximum capped at $2000$ units per transfer. \\ \cline{2-3} 
\multicolumn{1}{|l|}{} & Hadoop & Generated by geo-distributed data analytics over Facebook's inter-datacenter WAN (distribution mean of $20$ units per transfer). \\ \cline{2-3} 
\multicolumn{1}{|l|}{} & Cache-follower & Generated by geo-distributed cache applications over Facebook's inter-datacenter WAN (distribution mean of $20$ units per transfer). \\ \hline
\end{tabular}}
\end{table*}

\section{Evaluation} \label{evaluations}
We considered various topologies and transfer size distributions as shown in Table \ref{table_evaluations}. We selected two research topologies with given capacity information on edges from the Internet Topology Zoo \cite{zoo}. We could not use other commercial topologies as the exact connectivity and link capacity information were not publicly disclosed. We also considered multiple transfer volume distributions including synthetic (light-tailed and heavy-tailed) and real-world Facebook inter-datacenter traffic patterns (Hadoop and Cache-follower) \cite{social_inside}. Transfer arrival pattern was according to Poisson distribution with a rate of $\lambda$ per timeslot. For simplicity, we assumed an equal number of receivers for all bulk multicast transfers per experiment. We performed simulations and Mininet emulations to evaluate {\name}.

We compare {\name} with multiple baseline techniques and QuickCast \cite{quickcast} which also focuses on partitioning receivers into groups for improved completion times. We were unable to evaluate our work against another recent work called BDS \cite{overlay_hkust}, which has been developed by Baidu, due to source code unavailability. BDS takes advantage of store-and-forward and operates at the application layer. We qualitatively compare {\name} with BDS. {\name} can offer lower bandwidth consumption since BDS uses paths instead of trees, and can more effectively exercise physical links as it manages traffic at the network layer. On the other hand, since BDS uses all available overlay paths (possibly many routes to any specific receiver), under the lightly loaded regime, BDS may offer higher throughput (at the cost of considerably higher bandwidth consumption). Extension of {\name} to use parallel trees is considered part of future work.

\subsection{Computing a Lower Bound} \label{aggregate_topo}
\begin{figure}
    \centering
    \includegraphics[width=0.7\columnwidth]{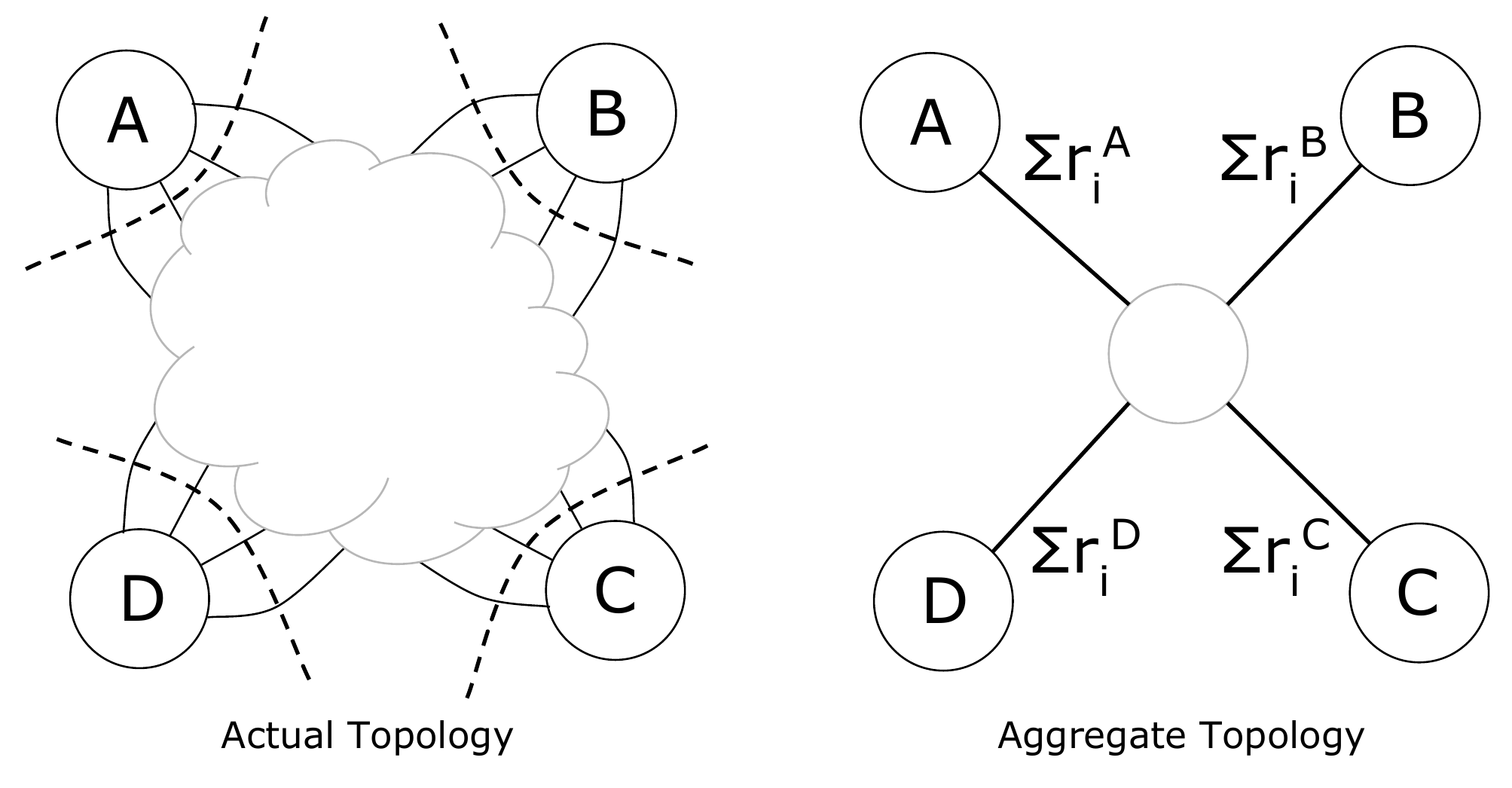}
    \caption{The physical topology, and the aggregate topology to compute a lower bound on receiver completion times. The aggregate topology is used only for evaluation purposes and it does not play any part in the design of \name.}
    \label{fig:aggregate_topology}
\end{figure}

We develop a technique to compute a lower bound on receiver completion times by creating an aggregate topology from the actual topology. As shown in Figure \ref{fig:aggregate_topology}, to create the aggregate topology, we combine all downlinks and uplinks with rates $r_i^{[node]}$ for all interfaces $i$ per node to a single uplink and downlink with their rates set to the sum of rates of physical links. Also, the aggregate topology connects all nodes in a star topology using their uplinks and downlinks and so assumes no bottlenecks within the network. Since this topology is a relaxed version of the physical topology, any solution that is valid for the physical topology is valid on this topology as well. Therefore, the solution to the aggregate topology is a lower bound that can be computed efficiently but may be inapplicable to the actual physical topology. We will use this approach in \S \ref{min_avg_comp} for evaluation of {\name}.

\subsection{Simulations}
In simulations, we focus on computing gains and therefore assume no dropped packets and accurate max-min fair rates. We normalized link capacities by maximum link rate per topology and fixed the timeslot length to $\delta = 1.0$.

\vspace{0.5em}
\noindent\textbf{Effect of User Traffic:} We account for the effect of higher priority user traffic in the simulations. The amount of available bandwidth per edge per timeslot, i.e., $B_e(t)$, is computed by deducting the rate of user traffic from the link capacity $C_e$. Recent work has shown that this rate can be safely estimated \cite{tempus, netstitcher}. For evaluations, we assume that user traffic can take up to $30\%$ of a link's capacity with a minimum of $5\%$ and that its rate follows a periodic pattern going from low to high and to low again. Per link, we consider a random period in the range of $10$ to $100$ timeslots that is generated and assigned per experiment instance.

\subsubsection{Minimizing Average Completion Times} \label{min_avg_comp}
This is when the objective vector is made of all ones. The partitioning hierarchy then begins with all receivers forming their $1$-receiver partitions. This is a highly general objective and can be considered as the default approach when the application/user does not specify an objective vector. We discuss multiple simulation experiments.

\begin{figure*}
    \centering
    \includegraphics[width=\textwidth]{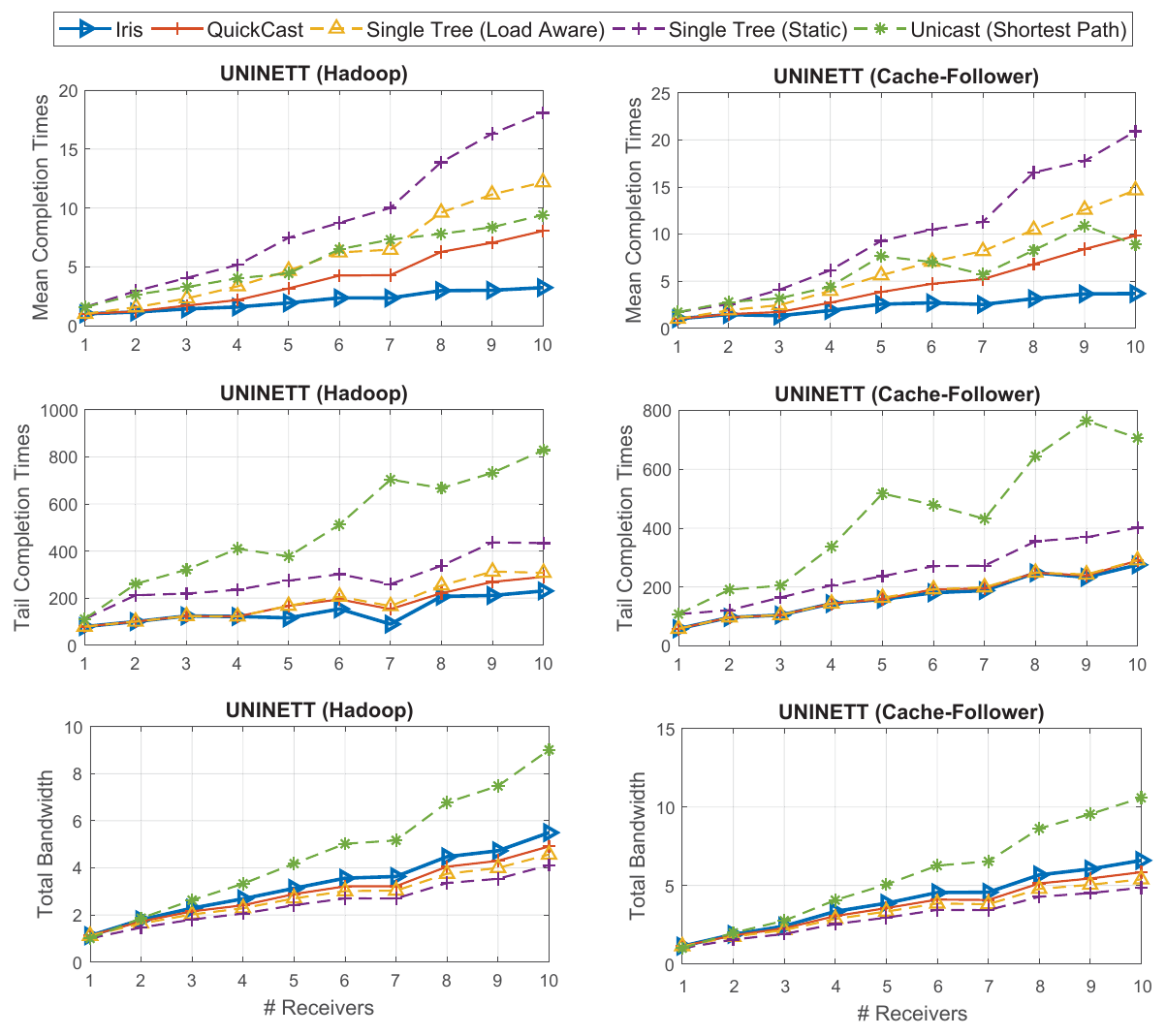}
    \caption{Comparison of various techniques by number of multicast receivers. Plots are normalized by the minimum data point (mean and tail charts are normalized by the same minimum), $\lambda = 1$, and lower values are better.}
    \label{fig:overall}
\end{figure*}

In Figure \ref{fig:overall}, we measure the completion times (mean and tail) as well as bandwidth consumption by the number of receivers (tail is 99.9\textsuperscript{th} percentile). We consider two baseline cases: unicast shortest path and static single tree (i.e., minimum edge Steiner tree) routing. The shortest path routing is the unicast scenario that uses minimum bandwidth possible. The minimum edge Steiner tree routing uses minimum bandwidth possible while connecting all receivers with a single tree. The first observation is that using unicast, although leads to highest separation of fast and slow receivers, does not lead to the fastest completion as it can lead to many shared bottlenecks and that is why we see long tail times. {\name} offers the minimum completion times (mean and tail) across all scenarios. Also, its completion times grow much slower compared to others as the number of receivers (and so overall network load) increases. This is while {\name} uses only up to $35\%$ additional bandwidth compared to the static single tree (unicast shortest path routing uses up to $2.25\times$). Compared to QuickCast, {\name} offers up to $26\%$ lower tail times and up to $2.72\times$ better mean times while using up to $13\%$ extra bandwidth.

\begin{figure*}
    \centering
    \includegraphics[width=0.8\textwidth]{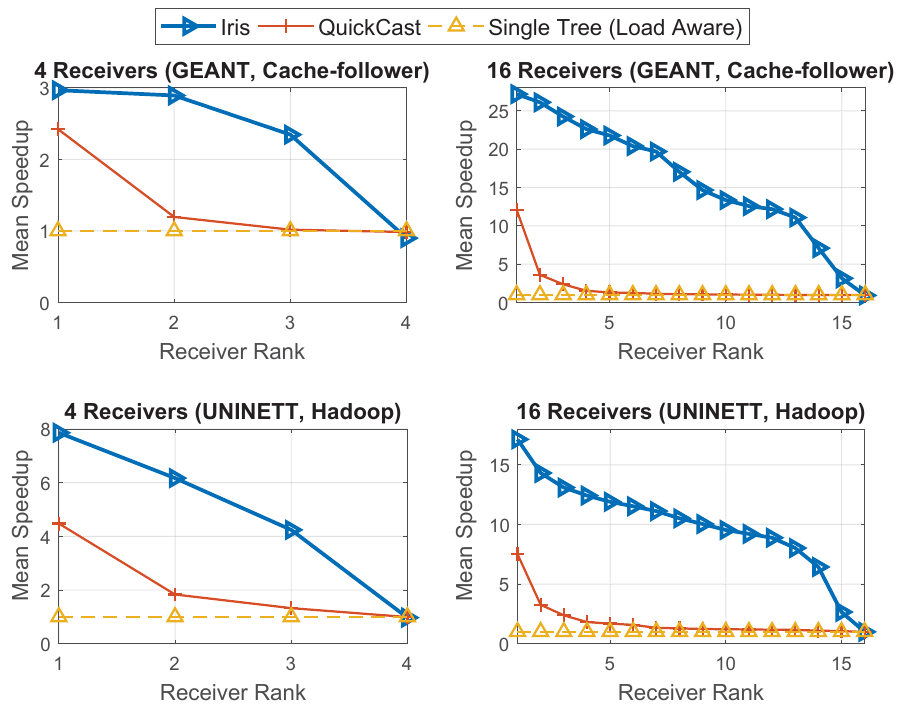}
    \caption{Mean completion time speedup (larger is better) of receivers normalized by no partitioning (load aware single tree) case given their rank from fastest to slowest, every node initiates equal number of transfers, receivers were selected according to uniform distribution from all nodes, and we considered $\lambda$ of 1.}
    \label{fig:speedup_1}
\end{figure*}

In Figure \ref{fig:speedup_1}, we show the completion times speedup of receivers by their rank. As seen, gains depend on the topology, traffic pattern, and receiver's rank. The dashed line is the baseline, i.e., no-partitioning case. Compared to QuickCast \cite{quickcast}, the fastest node always completes faster and up to $2.25\times$ faster with {\name}. Also, the majority of receivers complete significantly faster. In case of four receivers, the top $75\%$ receivers complete between $2\times$ to $4\times$ faster than baseline and with sixteen receivers, the top $75\%$ receivers complete at least $8\times$ faster than baseline. This is when QuickCast's gain drops quickly to one after the top $25\%$ of receivers.

\begin{figure*}
    \centering
    \includegraphics[width=0.8\textwidth]{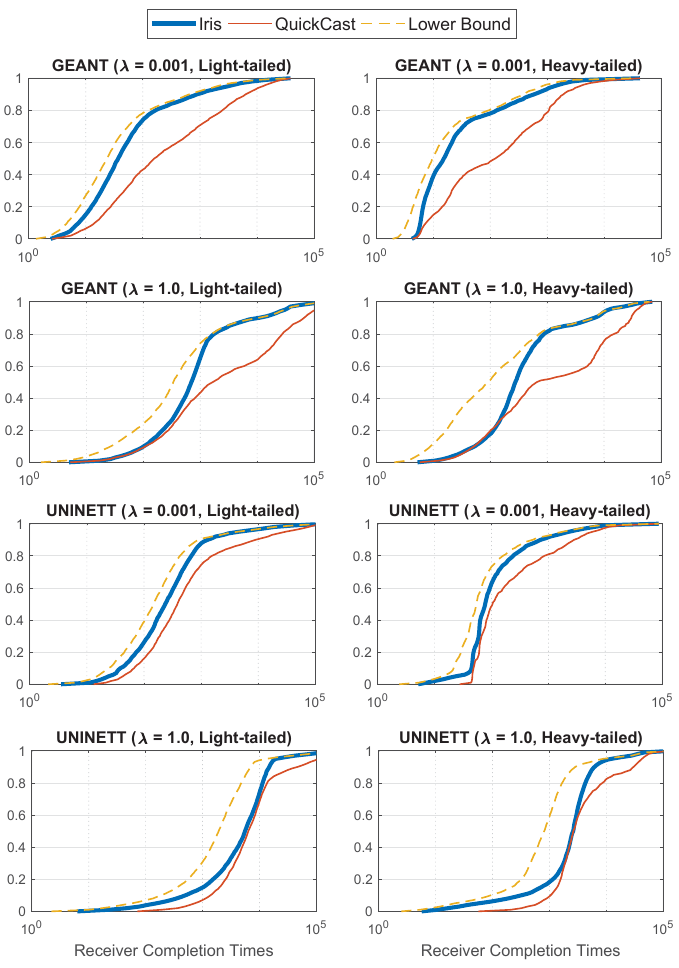}
    \caption{CDF of receiver completion times. Every transfer has 8 receivers selected uniformly across all nodes. ``Lower Bound'' is computed by finding the aggregate topology and applying Theorem 2.}
    \label{fig:speedup_2}
\end{figure*}

In Figure \ref{fig:speedup_2}, we measure the CDF of completion times for all receivers. As seen, tail completion times are two to three orders of magnitude longer than median completion times which is due to variable link capacity and transfer volumes. We evaluate the completion times of QuickCast and {\name} and compare them with a lower bound which considers the aggregate topology (see \S \ref{aggregate_topo}) and applies Theorem 2 directly. It is likely that no feasible solution exists that achieves this lower bound. Under low arrival rate (light load), we see that {\name} tracks the lower bound nicely with a marginal difference. Under high arrival rate (heavy load), {\name} stays close to the lower bound for lower and higher percentiles while not far from it for others.

\subsubsection{Other Objective Vectors}
\begin{figure*}
    \centering
    \includegraphics[width=0.9\textwidth]{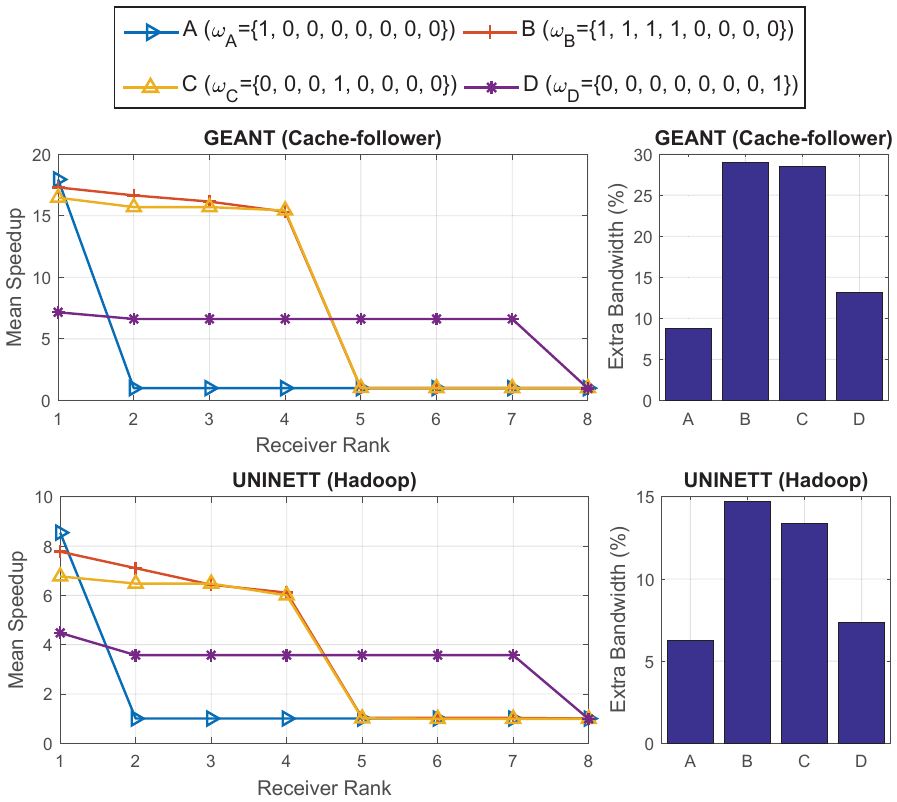}
    \caption{Gain by rank for different receivers per transfer averaged over all transfers for four different objective vectors. We set $\lambda = 0.1$ and there are $8$ receivers.}
    \label{fig:vectors}
\end{figure*}

We discuss four different objective vectors of $A$, $B$, $C$ and $D$ as shown in Figure \ref{fig:vectors}. This figure shows the mean speedup of receivers given their ranks, and the bandwidth consumption associated with each vector. In $A$, we aim to finish one copy quickly while not being concerned with completion times of other receivers. We see a gain of between $9\times$ to $18\times$ across the two topologies considered for the first receiver. We also see that this approach uses much less extra bandwidth compared to when we have a vector with more ones (e.g., case $B$). In $B$, we aim to speed up the first four receivers (we care about each one) while in $C$, we want to speed up the fourth receiver not directly concerning ourselves with the top three receivers. As can be seen, $B$ offers increasing speedups for the top three receivers while $C$'s speedup is flatter. Also, $C$ uses less bandwidth compared to $B$ by grouping the top three receivers into one partition at the base of the hierarchy. Finally, $D$'s vector specifies that the application/user only cares about the completion time of the last receiver which means that receiver will be put in a separate partition at the base of the hierarchy while other receivers will be grouped into one partition. Since the slowest receiver is usually limited by its downlink speed, this cannot improve its completion time. However, with minimum extra bandwidth, this speeds up all receivers except the slowest by as much as possible. Except for the slowest, all receivers observe a speedup of between $3\times$ to $6\times$ while using $8\%$ to $16\%$ less bandwidth compared to $B$. A tradeoff is observed, that is, $D$ offers lower speedup but consistent gain for more receivers with less bandwidth use compared to $B$.

\subsection{Mininet Emulations}
\begin{figure*}
    \centering
    \includegraphics[width=0.62\textwidth]{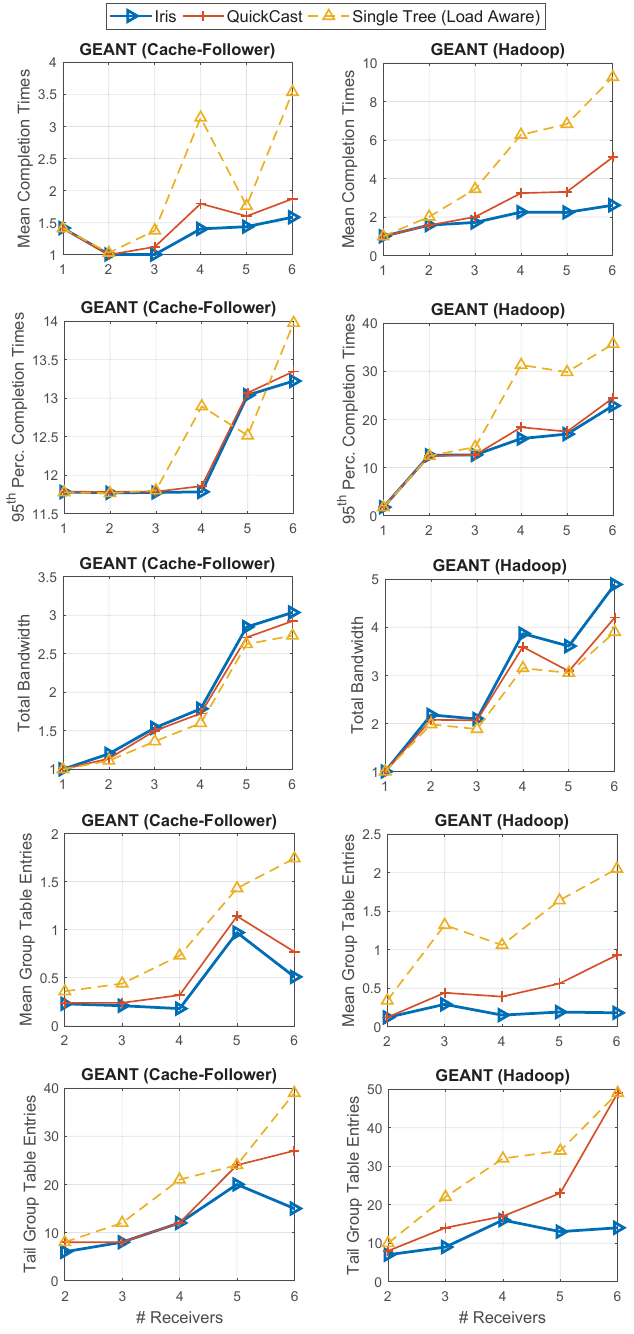}
    \caption{Mininet Emulation Results}
    \label{fig:mininet}
\end{figure*}

We used Mininet to build and test a prototype of {\name} and compare it with QuickCast and set up the testbed on CloudLab \cite{cloudlab}. We used OpenvSwitch (OVS) 2.9 in the OpenFlow 1.3 compatibility mode along with the Floodlight controller 1.2 connecting them to a control network. We assumed fixed available bandwidth over edges according to GEANT topology \cite{geant} while scaling downlinks' capacity so that the maximum is 500 Mbps.\footnote{In general, inter-datacenter link capacities may go beyond tens of Gbps. Due to the limitations of our emulation server, we had to use 500 Mbps as the maximum link capacity. That is, although we used a server with 56 logical CPU cores, even with a maximum link rate of 500 Mbps, the machine ran at close to full CPU utilization across most cores. Using a higher rate would have led to inaccurate emulation results due to the timing inaccuracies caused at high CPU utilization. The high CPU utilization in Mininet is caused mostly by the traffic shapers that Mininet uses to model links' capacities. Using a lower rate, however, should have a negligible effect on the validity of results as a proof of concept.} We did this to reduce the CPU overhead of traffic shaping over TCLink Mininet modules. Our traffic engineering program communicated with end-points through a RESTful API. We used NORM \cite{norm_navy} for multicast session management along with its rate-control module. To increase efficiency, we computed max-min fair rates centrally at the traffic engineering program and let the end-points shape their traffic using NORM's rate control module. The experiment was performed using twelve trace files generated according to Facebook traffic patterns concerning transfers' volumes \cite{social_inside}, and each trace file had 200 requests in total with an arrival rate of one request per timeslot based on Poisson distribution. We also considered timeslots of one second, a minimum transfer volume of 5 MB and limited the maximum transfer volume to 500 MB.\footnote{These parameters also match the distribution of YouTube video sizes \cite{you_tube}.} We considered three schemes of {\name}, QuickCast and a single tree approach (no partitioning). The total emulation time was about 24 hours. Figure \ref{fig:mininet} shows our emulation results. To allow comparison between the tail, i.e., 95\textsuperscript{th} percentile, and mean values, we have normalized both plots by the same minimum in each row. Also, the group table usage plots are not normalized and show the actual average and actual maximum across all switches. The reason why data points jump up and down is the randomness of generated traces that comes from transfers (volume, source, receivers, arrival pattern, etc).

\vspace{0.5em}
\noindent\textbf{Completion Times and Bandwidth:} {\name} offers up to $2.5\times$ speed up in mean completion times compared to QuickCast and $4\times$ compared to using a single multicast tree per transfer. We also see that compared to using a single multicast tree, {\name} consumes at most $25\%$ extra bandwidth.

\vspace{0.5em}
\noindent\textbf{Forwarding Plane:} We see that {\name} uses up to about $4\times$ less group table entries at the switches where the maximum number of entries were exhausted which allows more parallel transfers across the same network. {\name} achieves this by allowing a larger number of partitions per transfer whenever it does not hurt the completion times. By allowing more partitions, each tree will branch less times on average reducing the number of group table entries.

\vspace{0.5em}
\noindent\textbf{Running Time:} Across all experiments, the computation time needed to run {\name} to calculate partitions and forwarding trees, i.e., running Algorithm \ref{wancast_algorithm}, stayed below 5 ms per request.

\subsection{Practical Concerns}
New challenges, such as increased communication latency across network elements and failures, may arise while deploying {\name} on a real-world geographically distributed network. Communication latency may not affect the performance considerably as we focus on long-running internal transfers that are notably more resilient to latency overhead of scheduling and routing compared to interactive user traffic. Failures may affect physical links or the traffic engineering server. Loss of a physical link can be addressed by rerouting the affected transfers reactively either by the controller or by use of SDN fast failover mechanisms. End-points may be equipped with distributed congestion control (similar to \cite{mctcp}) which they can fall back to in case the centralized traffic engineering fails.

\section{Conclusions and Future Work}
Replication of content and data across geographically dispersed datacenters creates a large volume of multicast traffic that needs to be managed for increased performance. A bulk multicast transfer can be indicated with a source, set of receivers and total transfer volume. In this paper, we focused on the problem of grouping receivers into multiple partitions to minimize the effect of receiver downlink speed discrepancy on completion times of receivers. We analyzed a relaxed version of this problem and proposed a partitioning that reduces mean completion times of multicast receivers given max-min fair rates. We also set forth the idea of applications/users expressing their requirements in the form of binary objective vectors which allows us to optimize resource consumption and performance further. We then proposed {\name}, a system that computes partitions and forwarding trees for incoming bulk multicast transfers as they arrive given objective vectors. We showed that {\name} could significantly reduce mean completion times with a small increase in bandwidth consumption and can fulfill the requirements expressed using objective vectors while saving bandwidth whenever possible. It is worth noting that performance of any partitioning and forwarding tree selection algorithm rests profoundly on the network topology and transfer properties. Study of rate allocation policies besides max-min fairness and handling failures are among future directions.


\bibliographystyle{unsrt}
\bibliography{citations}

\end{document}